\documentclass[sigconf,screen]{acmart}
\renewcommand\footnotetextcopyrightpermission[1]{} 

\makeatletter
\renewcommand\@formatdoi[1]{\ignorespaces}
\makeatother

\usepackage{algorithm}
\usepackage{algorithmic}

\usepackage{microtype}

\usepackage{subcaption}
\usepackage{amsmath} 
\usepackage{nth}
\usepackage{xspace}

\AtBeginDocument{%
  }

\setcopyright{none}
\copyrightyear{2024}
\acmYear{2024}
\acmDOI{}
\acmISBN{}

\acmConference[RecSys FAccTRec Workshop '24]{7th FAccTRec Workshop: Responsible Recommendation}{Oct 14,
  2024}{Bari, Italy}

\begin{document}

\title{Reducing Population-level Inequality Can Improve Demographic Group Fairness: a Twitter Case Study}



\author{Avijit Ghosh}
\authornote{Work completed when author was at Northeastern University.}
\affiliation{%
  \institution{Hugging Face, University of Connecticut}
  \country{}
  }
  
\email{avijit@huggingface.co}

\author{Tomo Lazovich}
\authornotemark[1]
\affiliation{%
  \institution{U.S. Census Bureau}
  \country{} 
  }
\email{tomo.lazovich@census.gov}

\author{Kristian Lum}
\affiliation{%
  \institution{University of Chicago Data Science Institute}
  \country{} 
  }
\email{kristianlum@gmail.com}

\author{Christo Wilson}
\affiliation{%
  \institution{Northeastern University}
  \country{} 
  }
\email{cbw@ccs.neu.edu}

\renewcommand{\shortauthors}{Ghosh et al.}

\makeatletter
\newcommand{\ie}{\emph{i.e.}\@ifnextchar.{\!\@gobble}{}}
\newcommand{\eg}{\emph{e.g.}\@ifnextchar.{\!\@gobble}{}}
\newcommand{\etc}{etc\@ifnextchar.{}{.\@}}
\makeatother

\begin{abstract}
Many existing fairness metrics measure group-wise demographic disparities in system behavior or model performance. Calculating these metrics requires access to demographic information, which, in industrial settings, is often unavailable. By contrast, economic inequality metrics, such as the Gini coefficient, require no demographic data to measure. However, reductions in economic inequality do not necessarily correspond to reductions in demographic disparities. In this paper, we empirically explore the relationship between demographic-free inequality metrics-- such as the Gini coefficient-- and standard demographic bias metrics that measure group-wise model performance disparities specifically in the case of engagement inequality on Twitter. We analyze tweets from 174K users over the duration of 2021 and find that demographic-free impression inequality metrics are positively correlated with gender, race, and age disparities in the average case, and weakly (but still positively) correlated with demographic bias in the worst case. We therefore recommend inequality metrics as a potentially useful proxy measure of average group-wise disparities, especially in cases where such disparities cannot be measured directly. Based on these results, we believe they can be used as part of broader efforts to improve fairness between demographic groups in scenarios like content recommendation on social media.  
\end{abstract}



\maketitle

\section{Introduction}
\label{sec:intro}

The measurement of algorithmic bias in sociotechnical systems has gained attention from academics~\cite{buolamwini2018gender,sapiezynski2019algorithms}, the media~\cite{angwin2019machine,leon2021amazon}, and regulators~\cite{eureg,ukreg,usreg} in recent times. Due to the growing awareness about the harms of unfair algorithms, especially those used in critical decision-making processes~\cite{angwin2019machine,lemonadebias}, different measures of algorithmic fairness have been suggested by academia~\cite{mehrabi2019survey}, and sometimes lawmakers~\cite{us1979questions}. Research has also suggested debiasing methods~\cite{nam2020learning,liang2020towards} for algorithms using these metrics. 
	
Many of the existing notions of algorithmic ``bias'' are defined in terms of demographic disparities in a single model's performance~\cite{corbett2018measure,mehrabi2019survey,narayanan21fairness}. Solutions designed to ameliorate ``bias'' within this framework, however, are unrealistic in many settings for two reasons. First, the demographic information required to calculate demographic disparities is often unavailable. Collecting such information may be outright illegal in specific contexts, like insurance and lending~\cite{bogen2020awareness}. For large-scale systems, collecting or annotating demographic information for all users is logistically difficult~\cite{andrus2021measure}, expensive~\cite{annotationcosts}, and has potential adverse privacy implications~\cite{chang2021privacy}. An alternative approach to evaluating disparities in the absence of ground-truth demographic labels is inferring the demographic labels and evaluating model or system disparities conditional on the inferred labels. This approach requires great care to account for differential inaccuracy in the demographic label classifier itself~\cite{diana2021multiaccurate} as naive applications can, in fact, worsen the problem~\cite{ghosh2021fair}. Additionally, there are ethical issues with inferring sensitive attributes of non-consenting individuals~\cite{unethical-prediction}.
	
The second reason that the standard framework for ``bias" measurement and mitigation falls short is that it is tailored to machine learning models, not sociotechnical systems as a whole. Most proposed debiasing methods aim to add fairness constraints during the training of the machine learning algorithm~\cite{pmlr-v54-zafar17a,wanglabelnoise2021}. While these methods can reduce demographic disparities in model performance \textit{in situ}---i.e., in a lab setting---the additional computational cost rapidly adds up in a large sociotechnical system with multiple online models interacting with each other. For example, on Twitter\footnote{In 2023, Twitter was rebranded to be called X. All of the data in this analysis was collected from the Twitter platform in 2021 when it was still called Twitter, so we continue to refer to this analysis as pertaining to Twitter and Twitter data.} the decision to show a given tweet in a specific user's timeline is not a function of one model, but up to dozens of machine learning models and rule-based heuristics running in parallel and feeding into one another~\cite{knight2023}. Previous work has shown that ``fairness'' interventions on single models do not necessarily compose to ``fair'' outcomes at the sociotechnical system level~\cite{bhaskaruni2019improving,kenfack2021impact}. For these reasons, measurement and mitigation of sociotechnical system unfairness is challenging using most existing approaches. 

Recently, demographic-free approaches inspired by economic measures of inequality have been proposed~\cite{saint2020fairness,lazovich2021measuring}. These approaches measure skew in resource allocation rather than disparities in performance across demographic groups. Mitigation is achieved via experimentation to identify and avoid conditions where system output exceeds acceptable limits, rather than by tailoring loss functions to minimize between-group disparities. Evidence shows these methods can effectively mitigate unfairness: pymetrics builds an ensemble model by excluding models that fail the four-fifths rule~\cite{wilson2021building}, while LinkedIn reduces bias by avoiding experiment versions that increase overall inequality~\cite{saint2020fairness}.

Demographic-free, economic measures of sociotechnical system unfairness are conceptually appealing in their own right, as inequality in a system---however distributed across demographic groups---may be undesirable. There is no guarantee, however, that reducing inequality in the aggregate leads to reductions in demographic disparities, a phenomenon we illustrate in Section~\ref{sec:correspondence}. Given the importance of ensuring system errors and benefits are not disproportionately concentrated within demographic groups, we must ensure that interventions in the name of reducing inequality in general directly translate to reductions in inequality between demographic groups as well.

In this paper, we empirically investigate the relationship between demographic-free measures of inequality and demographic disparity metrics, focusing on engagement inequality on Twitter. Our research questions are:

\begin{itemize}
   \item Do demographic-free measures of engagement inequality correlate with demographic disparity metrics on Twitter?
   \item Does the correlation vary across demographic and sensitive attributes?
   \item How does the correlation differ between marginal and intersectional bias metrics?
\end{itemize}

To address these questions, we analyze a dataset of 269M tweets from 739K users linked to US voter registration records~\cite{hughes2021}. We examine tweet engagements (likes and retweets, as has been done in past work~\cite{engagement}) stratified by gender, race, and age, observing whether demographic-aware and demographic-free inequality metrics for engagements increase or decrease together.

Through this work, we make several key contributions. We provide a comprehensive analysis of correlations between demographic-free inequality metrics and demographic disparity metrics in a large-scale, real-world social media platform. Our findings reveal a correspondence between demographic disparities and demographic-free engagement inequality on Twitter, with Spearman's rank correlation coefficients up to 0.78. This suggests that using inequality metrics for system tuning could lead to substantial reductions in demographic disparities. We offer insights into how these correlations vary across different demographic attributes and their intersections, and discuss the implications and potential risks of using demographic-free measures as proxies in settings where demographic data is limited or unavailable. Our results indicate that decision-makers at Twitter could implement strategies to avoid deploying changes that worsen distributional inequality, offering a promising path to reduce demographic disparities in sociotechnical systems where demographic information is limited.

The remainder of this paper is structured as follows: Section~\ref{sec:background} provides background on inequality and disparity metrics. Section~\ref{sec:data} describes our dataset. Section~\ref{sec:methodology} outlines our methodology. We present our results in Section~\ref{sec:results}, discuss limitations in Section~\ref{sec:limitations}, and conclude with future work in Section~\ref{sec:conclusion}.

\begin{figure*}[t]
\centering
\includegraphics[width = 0.8\textwidth]{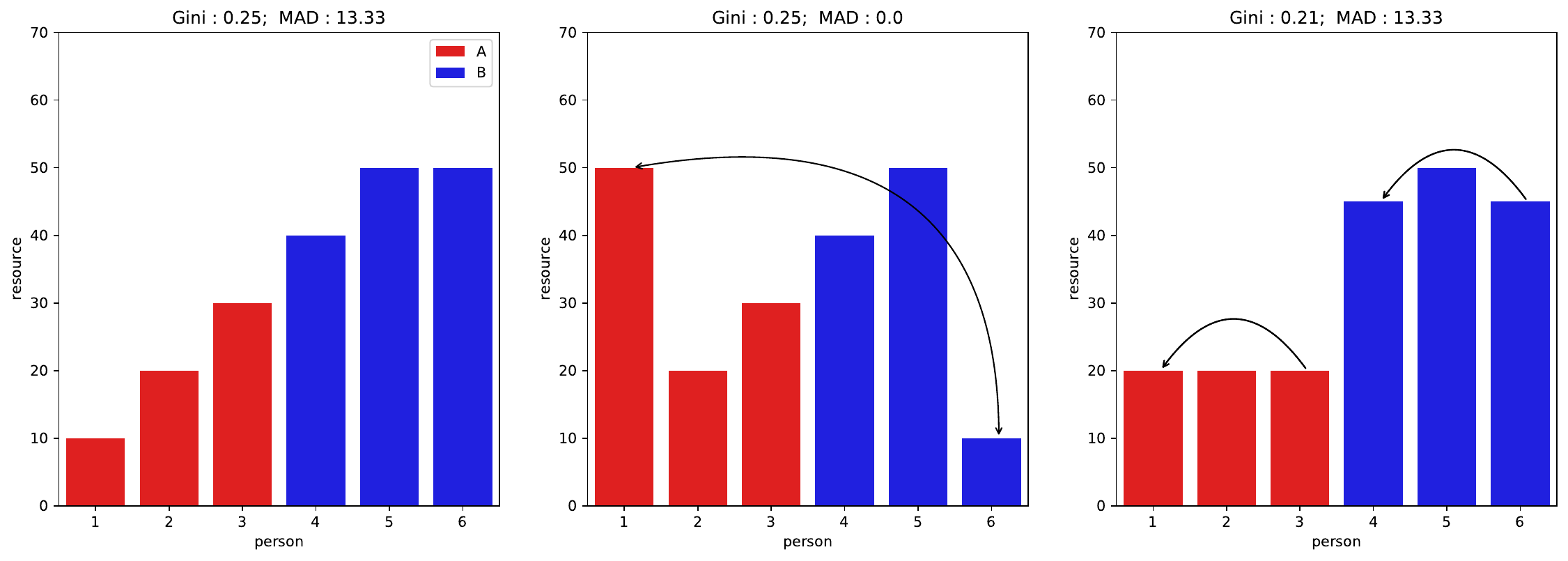}
\caption{Toy example illustrating that changes to Gini need not correspond to changes in demographic disparities and vice versa.}
\label{fig:gini-example}
\end{figure*}

\section{Background}
\label{sec:background}




This work analyzes the relationship between engagement inequality and demographic disparities in engagements on Twitter. Twitter is a social media platform in which users communicate with one another by authoring tweets---which may include short snippets of text, images, videos, links, etc. Users follow other users to receive the followees' tweets in their feeds. For example, if user $A$ follows user $B$, then the tweets that user $B$ authors may show up in user $A$'s feed. Users can choose to have tweets displayed in their feed via a curation algorithm in which Twitter puts the tweets it estimates to be most relevant to that user closer to the top, possibly including tweets from accounts the user does not follow. Or, users may select a reverse chronological algorithm, in which tweets authored by accounts the user follows are ordered such that the most recent such tweets appear at the top. Omitting some technical details, an impression occurs when a user views an author's tweet. 

Engagements are the currency of social media. Authors who receive many engagements and impressions are able to transmit their messages, opinions, and voice to many different people~\cite{tufekci2017twitter,meyer2021social}. While this in and of itself is valuable to some users, receiving many engagements can also translate more directly to other forms of wealth. For example, accounts with many followers (and thus many impressions) can monetize their accounts by being paid for using their platform to advertise products~\cite{kopf2020rewarding,goanta2023content}.

In the interest of making Twitter a place where a larger and more diverse set of individuals can utilize the platform to influence the world, have their voice heard, and potentially reap the economic benefits of the platform, it is important that tweet engagements are distributed equitably among Twitter users. To ensure this is happening, we must be able to measure inequality and demographic disparities on the platform. 


\subsection{Notation}

Consider first a general setting in which we are interested in differences in engagements across individuals or groups. In terms of notation, we let $I_j$ be the number of engagements individual $j$ receives on the platform. For demographic groups $k \in \mathcal{G}$, let $G_k$ be the set of indices, $j$, corresponding to individuals who belong to demographic group $k$. For examples, if groups are defined by age,  with age groups $\mathcal{G} = \{<18, 18-25, 26-45, 46-65, 66-85, 85+\}$, then $G_{k}$ for $k = `18-25'$ is a set consisting of the indices of all individuals between 18 and 25 years of age. We define the average engagements for group $k$ as
\begin{equation}
\bar{I}_{G_k} = \frac{\sum_{j : j \in G_k} I_j}{|G_k|}.
\end{equation}


\subsection{Measures of Demographic Disparities}

For each group $k \in \mathcal{G}$, we have $\bar{I}_{G_k}$, the average number of engagements received by individuals who belong to group $G$. When $|\mathcal{G}|$ is large this can be high dimensional, and we need a simple summary measure of variability across groups to quantify how differently average engagements are distributed across groups. We consider two such measures.

\subsubsection{Mean Absolute Deviation (MAD)} The Mean Absolute Deviation is defined as the average deviation of the impressions received per group from the average engagements over the entire distribution. A value of zero indicates that each group receives the exact same number of engagements. Higher values indicate larger disparities across groups. It is, therefore, a measure of the \textit{average case} of demographic bias. It is defined as

\begin{equation}
    \text{MAD} = \frac{\sum_{k \in \mathcal{G}} \left |\bar{I}_{G_k} - \frac{\sum \bar{I}_{G_k}}{|\mathcal{G}|}\right |}{|\mathcal{G}|}.
\end{equation}

\subsubsection{Inverse Min/Max (IMM)} Inverse Min/Max is defined as the minimum engagements for a particular subgroup over the maximum engagements for a particular subgroup, subtracted from 1. A value of zero indicates that the group with the minimum number of average engagements has the same number of average engagements as the group with the maximum number of average engagements, \ie the fairest scenario. IMM attempts to measure the \textit{worst case} of the demographic bias. It is defined as

\begin{equation}
    \text{IMM} = 1 - \frac{\min_{k \in \mathcal{G}} \bar{I}_{G_k}}{\max_{k \in \mathcal{G}} \bar{I}_{G_k}}.
\end{equation}

Several other measures for summarizing group-wise disparities have been proposed in recent work~\cite{lum2022biasing}. Here, we focus on just MAD and IMM, as these cover the two general categories of metrics for summarizing group-wise disparities~\cite{mathworksExploreFairness}---those that measure average differences, such as Statistical Parity Difference and Equal Opportunity Difference and those that look at the extremes, such as Disparate Impact.

\begin{figure*}[t]
    \centering

    \begin{subfigure}[t]{0.22\textwidth}
        \vtop{\includegraphics[width=\linewidth]{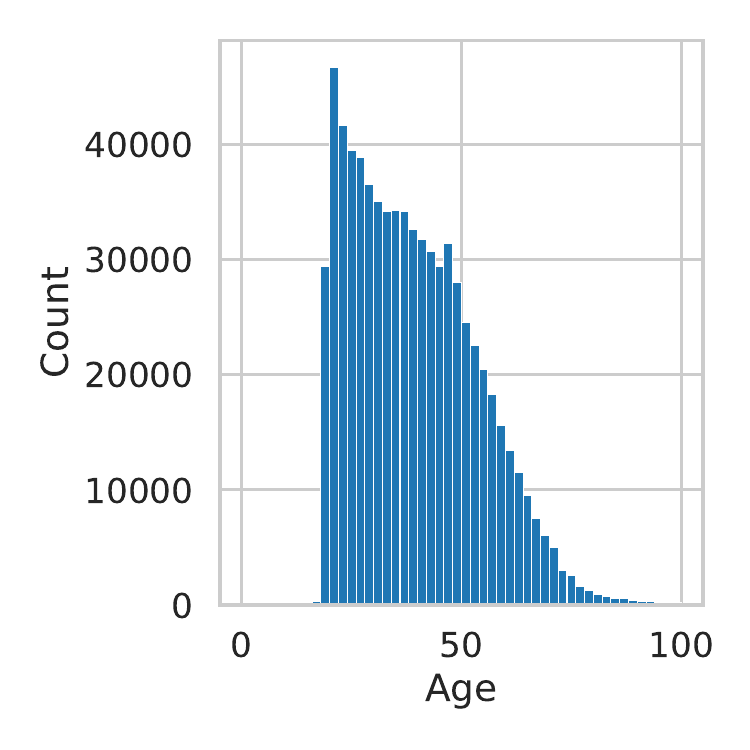}}
        \caption{}
        \label{fig:demo1}
    \end{subfigure}%
    \hfill
    \begin{subfigure}[t]{0.25\textwidth}
        \raisebox{1.9mm}{
        \vtop{\includegraphics[width=\linewidth]{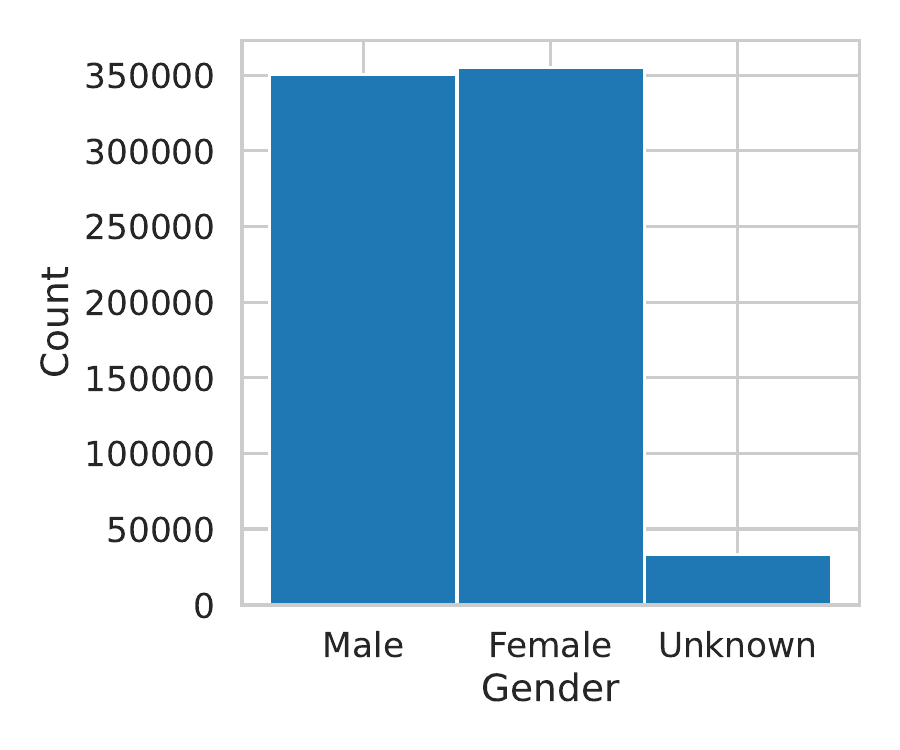}}}
        \caption{}
        \label{fig:demo2}
    \end{subfigure}%
    \hfill
    \begin{subfigure}[t]{0.22\textwidth}
        \vtop{\includegraphics[width=\linewidth]{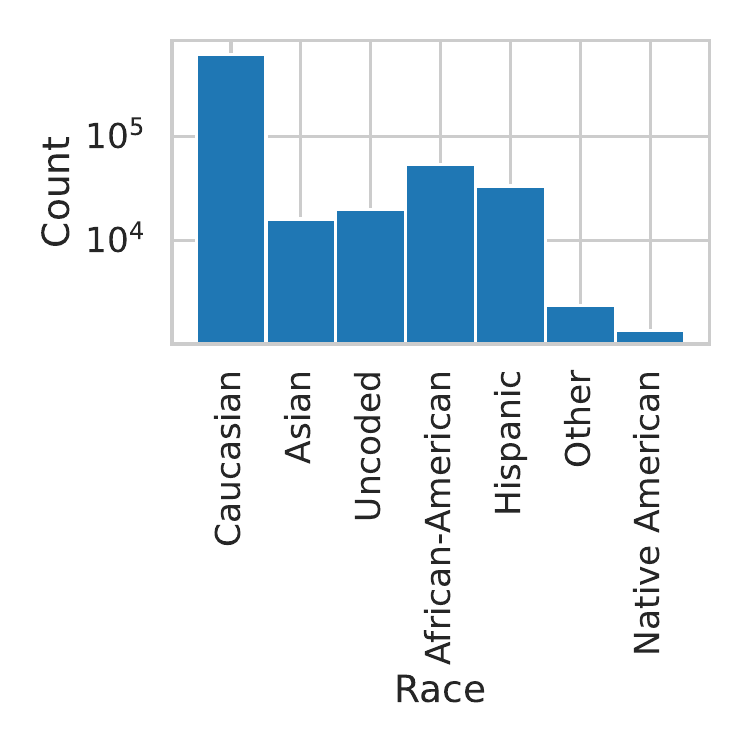}}
        \caption{}
        \label{fig:demo3}
    \end{subfigure}
    \hfill
    \begin{subfigure}[t]{0.22\textwidth}
        \vtop{\includegraphics[width=\linewidth]{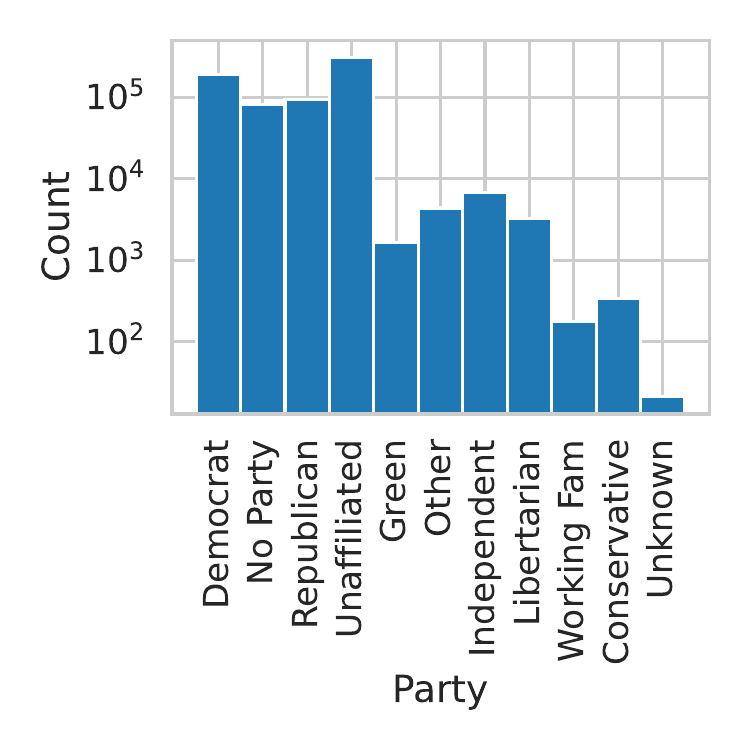}}
        \caption{}
        \label{fig:demo4}
    \end{subfigure}%

    \caption{Demographic distribution of the tweet authors in our dataset.}
    \label{fig:demo}
\end{figure*}

\subsection{Demographic-Free Inequality Measures}

Previous work has investigated economic inequality measures in the context of both experimentation and engagements~\cite{saint2020fairness,lazovich2021measuring}. For this work, we choose two inequality metrics that do not require demographic information: Gini coefficient and top 1\% share (which we refer to as T1PS going forward).

\subsubsection{Gini Coefficient} For a distribution of interest, the Gini coefficient is a measurement of the ratio of the average absolute difference between members of a population to the mean for that population~\cite{gini1912variabilita,farris2010gini}, defined as
\begin{equation}
    \text{Gini} = \frac{\sum_{p=1}^{N}\sum_{q=1}^{N}\left|I_p - I_q\right|}{2N\sum_{j=1}^{N} I_j}
\end{equation}
where $N$ is the total number of individuals in the population.

\subsubsection{Top 1\% Share (T1PS)} In everyday discussion of economics, it is often commonplace to hear statements such as ``the top X\% of people have Y\% of all wealth". Previous work has found that on Twitter, where the distribution of engagements is quite skewed, the top 1\% share of engagements specifically is a useful measure of inequality on the platform~\cite{lazovich2021measuring}. Let $T_{99}$ be the set of indices such that the user's value of engagements $I_j$ is greater than or equal to the \nth{99} percentile of the distribution. Then, the top 1\% share is defined as

\begin{equation}
    \text{T1PS} = \frac{\sum_{j\in T_{99}} I_j}{\sum_{p=1}^{N} I_p}.
\end{equation}

\subsection{Correspondence} \label{sec:correspondence}

Although intuition would suggest some correspondence between demographic-free measures of inequality and demographic disparities, this need not be the case. Consider the toy example illustrated in Figure~\ref{fig:gini-example}. Each bar corresponds to the allocation of a resource (perhaps, engagements) for individuals one through six, whose index is noted on the horizontal axis. The group to which each individual belongs is denoted by the color of the bar. In the left panel, the Gini coefficient is 0.25 with a MAD of 13.33.

Consider what happens if we swap person one and six's resources. Clearly, the overall inequality has not changed---the Gini coefficient in the middle panel is simply the Gini coefficient from the first panel with the indices for person one and six switched. However, MAD has been reduced to zero, with individuals in both groups receiving, on average, $\frac{100}{3}$ units of the resource. This shows that it is possible that demographic inequality can be changed without impacting demographic-free measures of inequality at all.
 
On the flip side, consider the right panel of Figure~\ref{fig:gini-example}. In this case, we again assume a starting point shown in the left panel. Now, person three transfers 10 units of resource to person one and person six transfers five units of resource to person four. In this case, the Gini coefficient drops from 0.25 to 0.21. However, because resource transfer has taken place only among individuals within groups, the average resource within groups has not changed and thus MAD has not changed. Therefore, it is also possible to observe reductions to demographic-free inequality measures while demographic disparities have not improved at all. 
 
These examples illustrate that there is no guarantee that tracking demographic-free measures of inequality will necessarily give us information about demographic disparities. However, whether re-allocation happens in ways that impact demographic-free metrics while leaving demographic disparities unchanged (or vice-versa) in real systems is an empirical question that we address for the case of engagements on Twitter below.

\begin{figure*}[htb!]
    \centering

    \begin{subfigure}{0.5\textwidth}
        \vtop{\includegraphics[width=\linewidth]
        {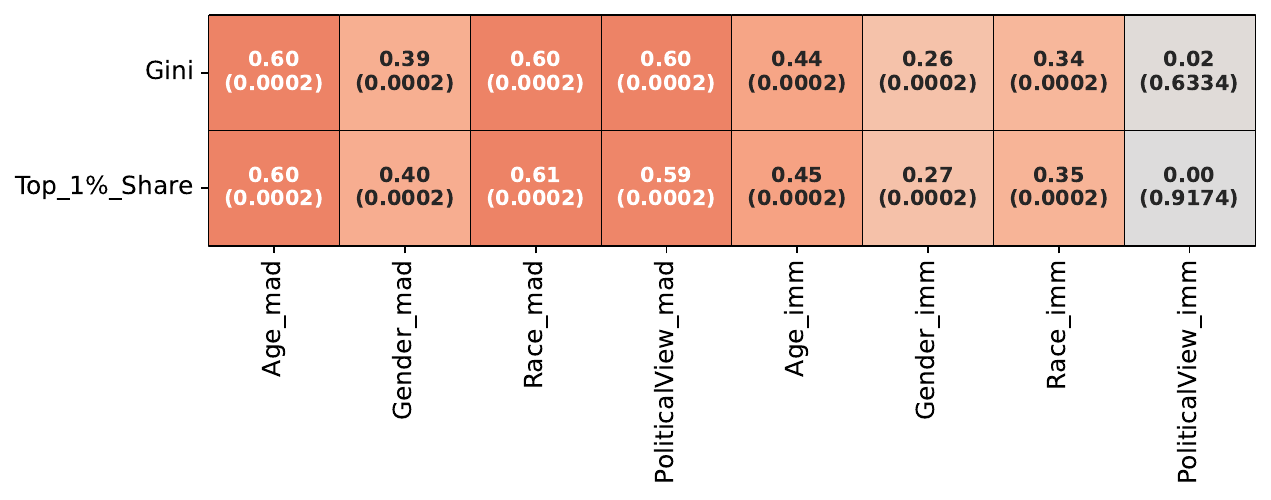}}
        \caption{Inequality metrics versus marginal bias metrics.}
        \label{fig:marginalsp}
    \end{subfigure}%
    \hfill
    \begin{subfigure}{0.75\textwidth}
        \vtop{\includegraphics[width=\linewidth]
        {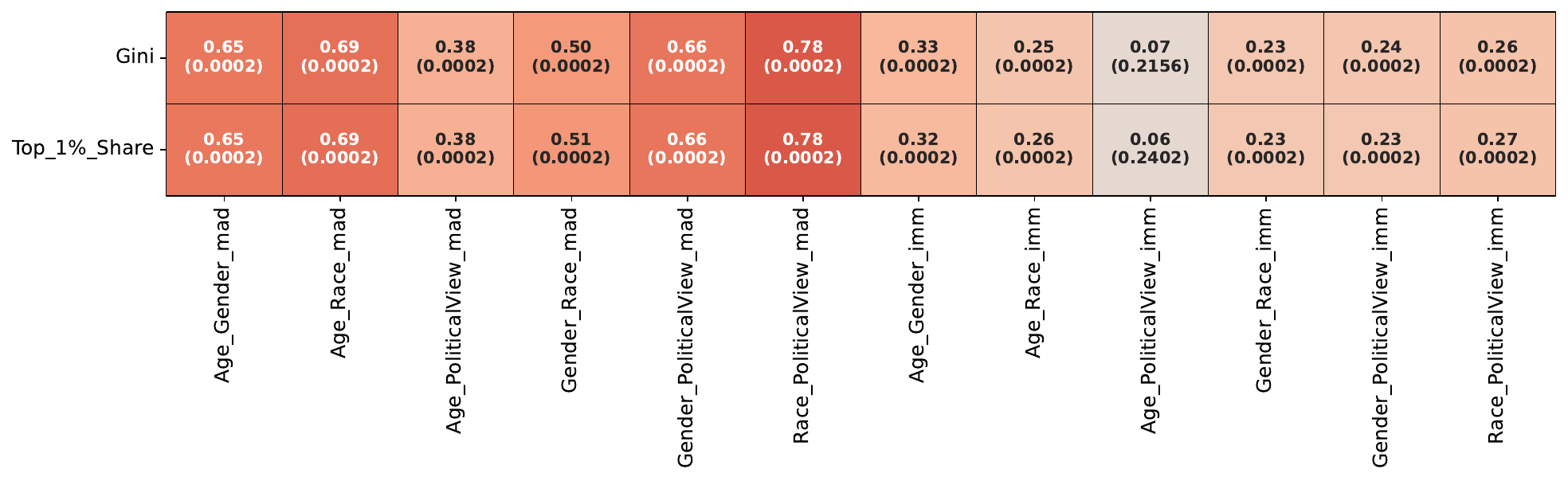}}
        \caption{Inequality metrics versus intersectional bias metrics.}
        \label{fig:intsp}
    \end{subfigure}

    \caption{Spearman's correlation and p-values between marginal and intersectional bias metrics. The two rows of each subfigure correspond to Gini coefficient and T1PS, respectively. The columns are the demographic bias metrics, either marginal or intersectional (combination of two marginal metrics), with the suffix denoting whether it was a MAD or IMM metric.}
    \label{fig:spearman_main}
\end{figure*}

\section{Dataset}
\label{sec:data}
	

For this experiment, we leveraged a dataset of 1.6 million Twitter users whose accounts were linked to public voter records provided by data vendor TargetSmart. The process of matching Twitter accounts to voter records, as well as in-depth demographic characteristics of the whole dataset, are described in detail in~\citet{hughes2021}. This dataset has been used in past work to characterize the spread of fake news on Twitter~\cite{grinberg2019} and public discourse during the first nine months of the COVID-19 pandemic~\cite{shugars2021}. This dataset was collected under Northeastern IRB protocol \#17-12-13.

Figure~\ref{fig:demo} shows the distributions of the different demographic attributes of the Twitter users in our dataset. We note the important caveat that in this dataset, when compared with a sample from Pew Research Center of Twitter users, female users are slightly over-represented, Hispanic users are slightly under-represented, and Asian users are significantly underrepresented~\cite{hughes2021}. Another limitation of the annotations provided is that gender values are limited to ``Male,'' ``Female,'' or ``Unknown,'' and the racial groups are limited to ``Caucasian,'' ``Asian,'' ``African American,'' ``Hispanic,'' ``Native American,'' ``Other,'' or ``Uncoded'' (meaning the race is not known). Further, we group the Age variable into discrete groups: $<18, 18-25, 26-45, 46-65, 66-85, \text{and } 85+$ respectively. Party affiliation (Political View) is annotated as tracked by the individual state the user is from and is one of ``Conservative,'' ``Democrat,'' ``Republican,'' ``Green,'' ``Libertarian,'' ``Independent,'' ``No Party,'' ``Other,'' ``Unaffiliated,'' or ``Unknown.''

For this work, we restrict the dataset to authors who have authored a tweet in the year 2021, and likewise only consider tweets from that year. We focus on tweet interactions, specifically likes and retweets, as our primary metric. For each tweet, we sum the number of likes and retweets to form a single metric called \textit{engagements}, following similar approaches in previous studies \cite{engagement}. Our choice to study engagements rather than impressions is based on both conceptual advantages and data availability constraints. Likes and retweets are publicly visible for the accounts in our dataset, whereas impressions are not. Moreover, engagements occur when a user is served content and chooses to interact with it, thus incorporating user feedback about content quality. This prevents artificial reduction of inequality by increasing low-quality content exposure to underrepresented groups. Additionally, our preliminary investigations showed that likes and retweets were positively correlated, and a disaggregated analysis would have revealed similar patterns. This combined engagement metric simplifies our analysis while still capturing the essential dynamics of user interaction with tweets.

For uniformity and removal of potential spurious trends, we only consider authors who had tweeted at least once every month. We also remove all tweet IDs that were retweets without quote, as the engagements on these tweets were added to the original tweet. This filtering reduced our dataset to 269M tweets from 174.6K unique authors.

\section{Methodology}
\label{sec:methodology}

To understand whether demographic-free inequality measures correlate with demographic disparity metrics, we do the following time series correlation analysis: 

\begin{enumerate}
    \item Compute inequality metrics for each day across the distribution of per-author engagements received for that day. 
    \item Compute the mean number of engagements by demographic group for each day. 
    \item Compute demographic bias metrics (MAD and IMM) for each day, using the mean number of engagements per group. We measure both marginal bias metrics and intersectional bias metrics. For intersectional bias, we limit ourselves to Cartesian products of pairs of marginal attributes, such as race $\times$ gender, age $\times$ political view, etc., as has been done in the literature~\cite{ghosh2021characterizing}.
    \item Measure Spearman's rank correlation coefficient~\cite{spearman1904proof} between the per day inequality metrics and demographic bias metrics, testing for joint monotonicity of the metrics. We used the permutation approach to calculate robust statistical significance values.\footnote{According to the Scipy stats guide on \url{https://docs.scipy.org/doc/scipy/reference/generated/scipy.stats.spearmanr.html}, it is recommended to use the permutation approach for correlations between small series, especially when we suspect autocorrelation and cannot be sure that the null hypothesis is an independent and identically distributed (IID) pair.}
\end{enumerate}




\begin{figure*}[t]
    \centering
    \begin{subfigure}[t]{0.24\textwidth}
        \includegraphics[width=\linewidth]{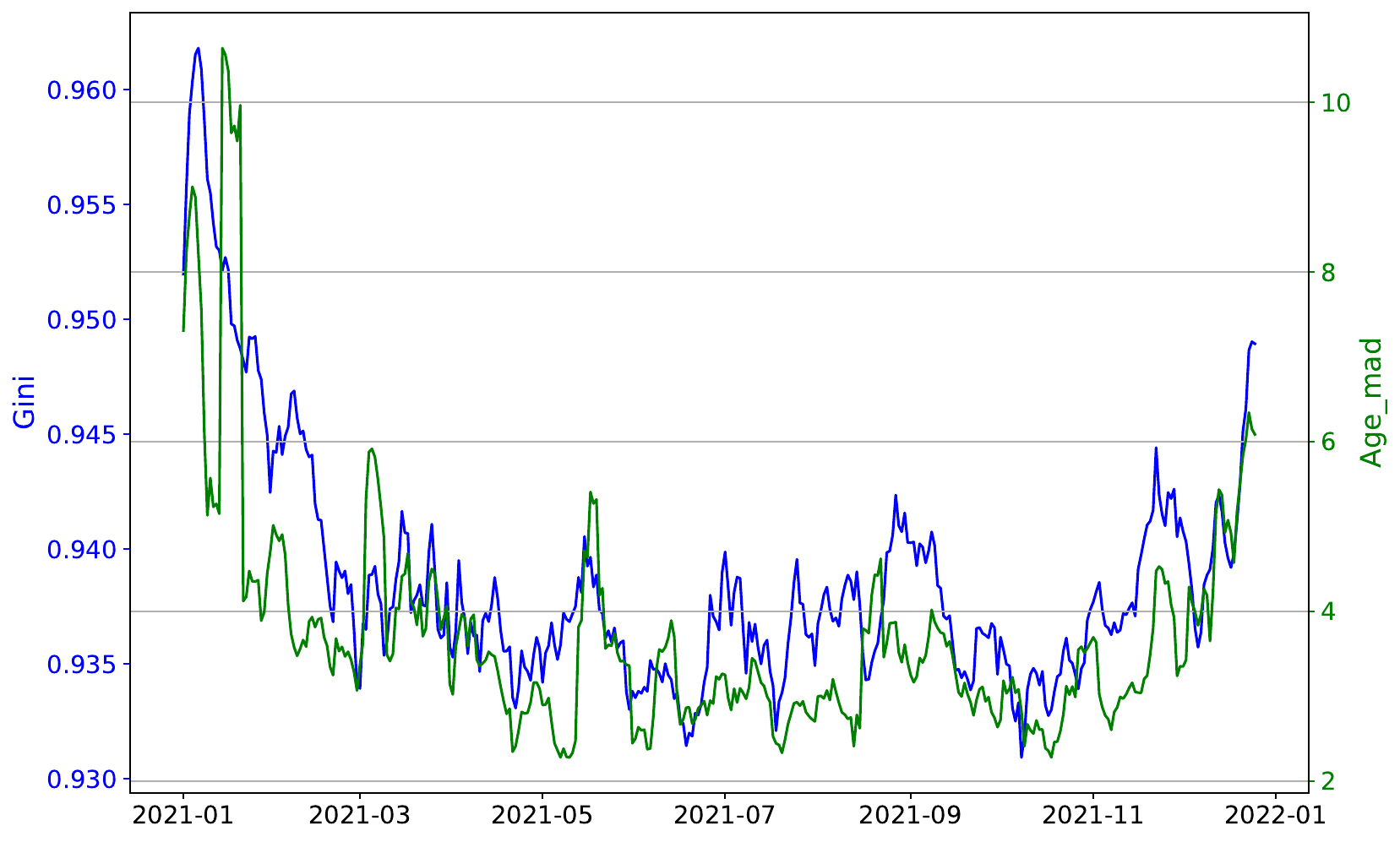}
        \caption{GINI vs. Age (MAD)}
        \label{fig:tt1a}
    \end{subfigure}%
    \hfill
    \begin{subfigure}[t]{0.24\textwidth}
        \includegraphics[width=\linewidth]{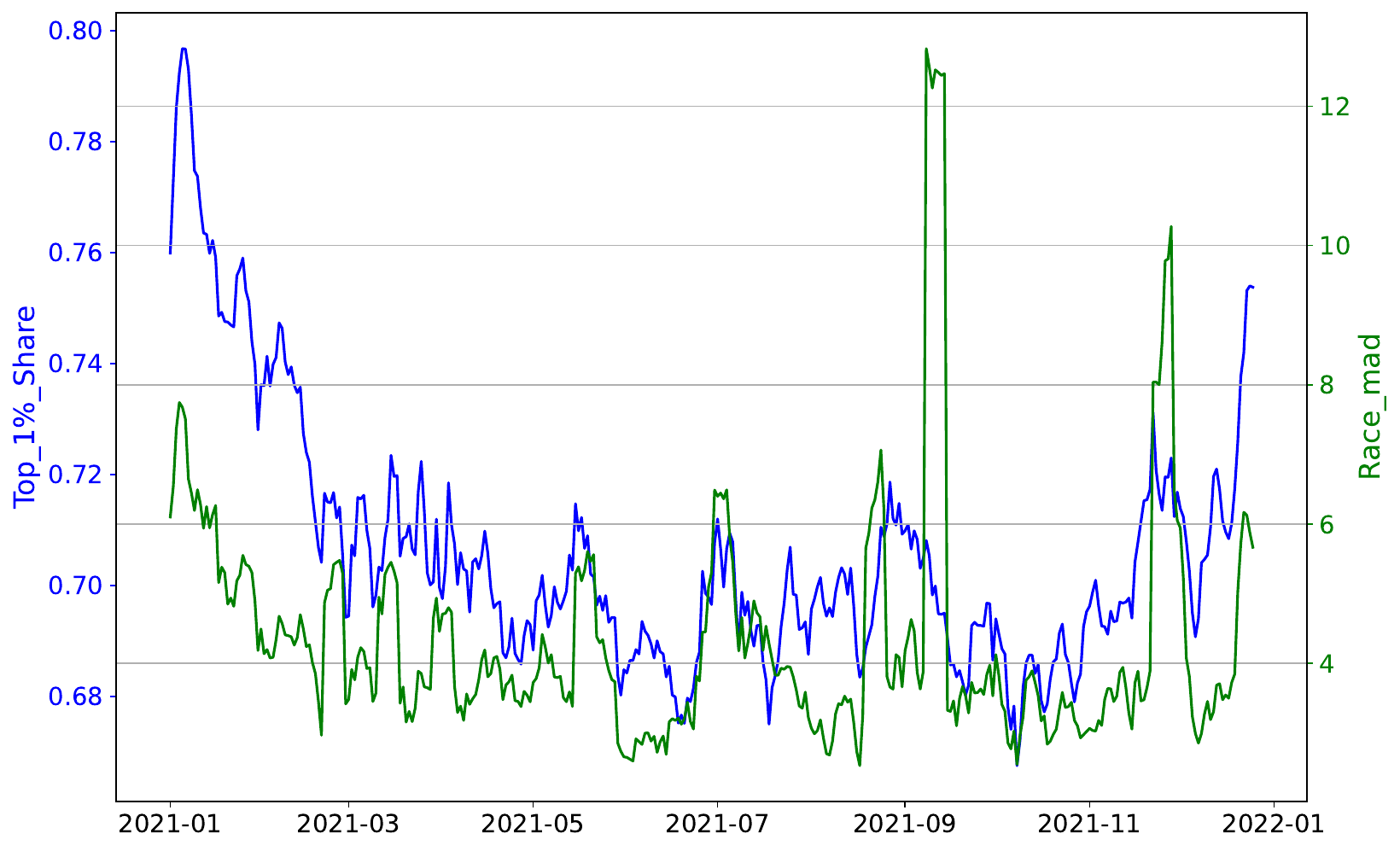}
        \caption{T1PS vs. Race (MAD)}
        \label{fig:tt1b}
    \end{subfigure}
    \hfill
    \begin{subfigure}[t]{0.24\textwidth}
        \includegraphics[width=\linewidth]{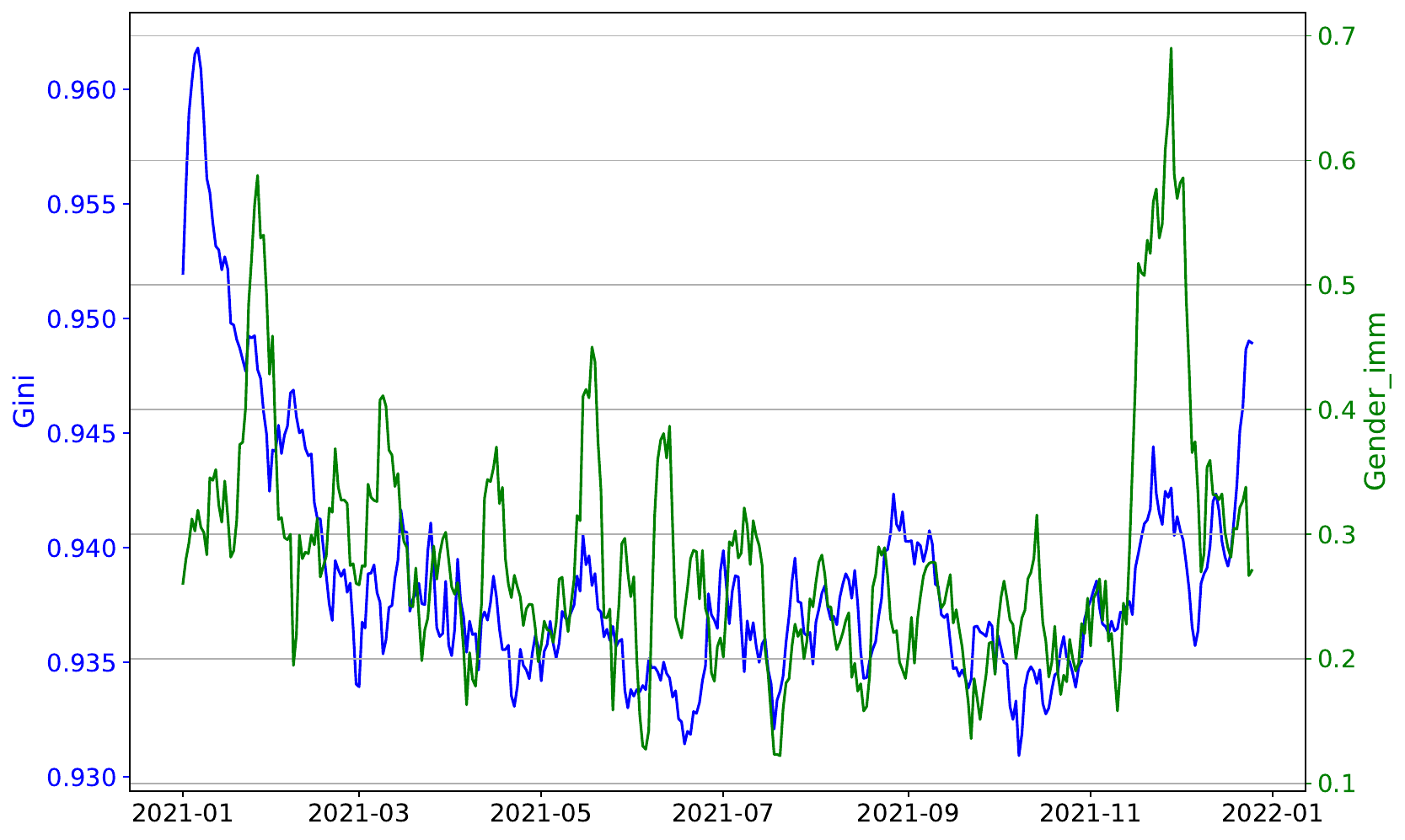}
        \caption{GINI vs. Gender (IMM)}
        \label{fig:tt1c}
    \end{subfigure}
    \hfill
    \begin{subfigure}[t]{0.24\textwidth}
        \includegraphics[width=\linewidth]{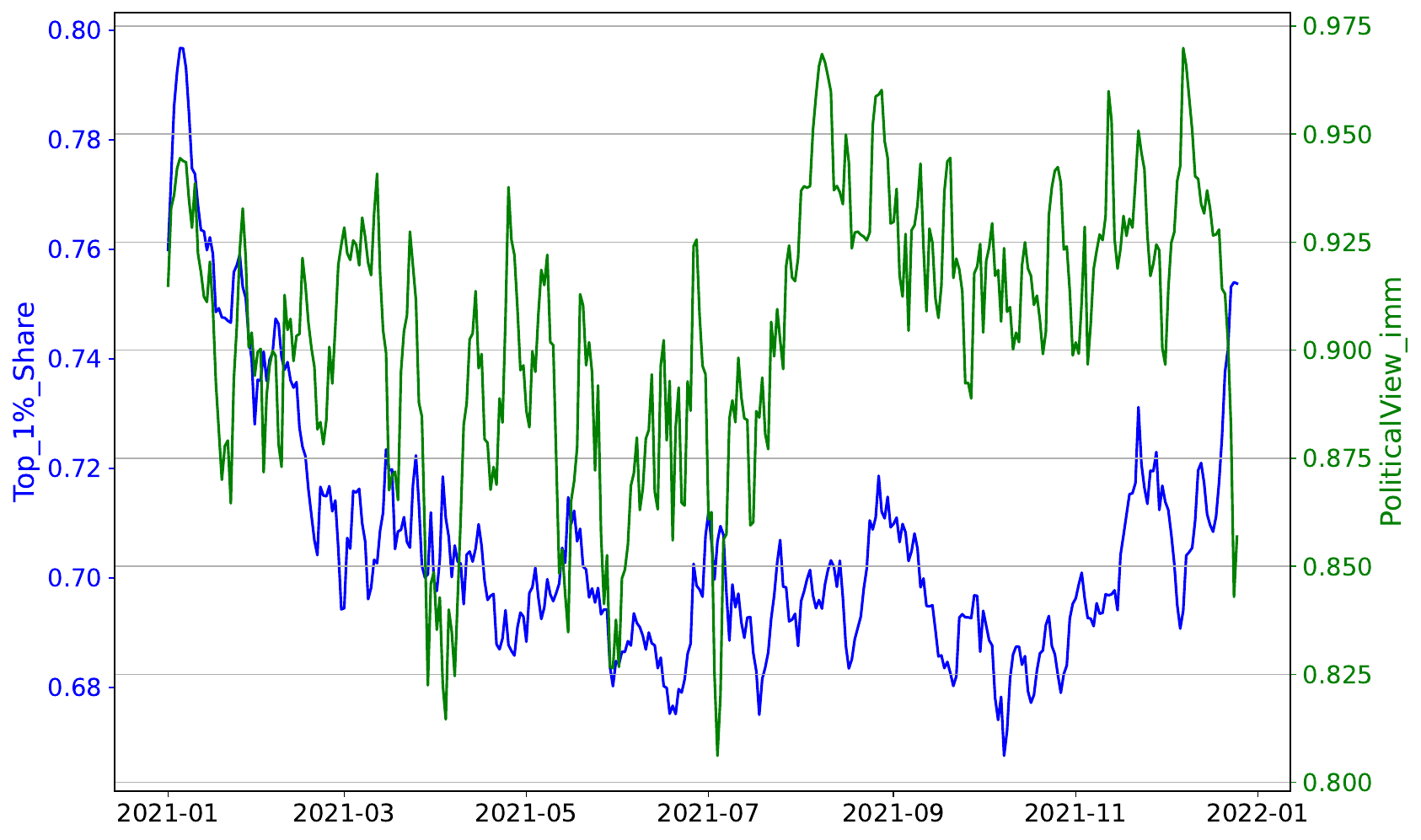}
        \caption{T1PS vs. Political View (IMM)}
        \label{fig:tt1d}
    \end{subfigure}
    \caption{Daily tracking of Inequality Metrics (blue) and Marginal Bias Metrics (green) over 2021.}
    \label{fig:timeseriesmarg}
\end{figure*}

\begin{figure*}[t]
    \centering
    \begin{subfigure}[t]{0.24\textwidth}
        \includegraphics[width=\linewidth]{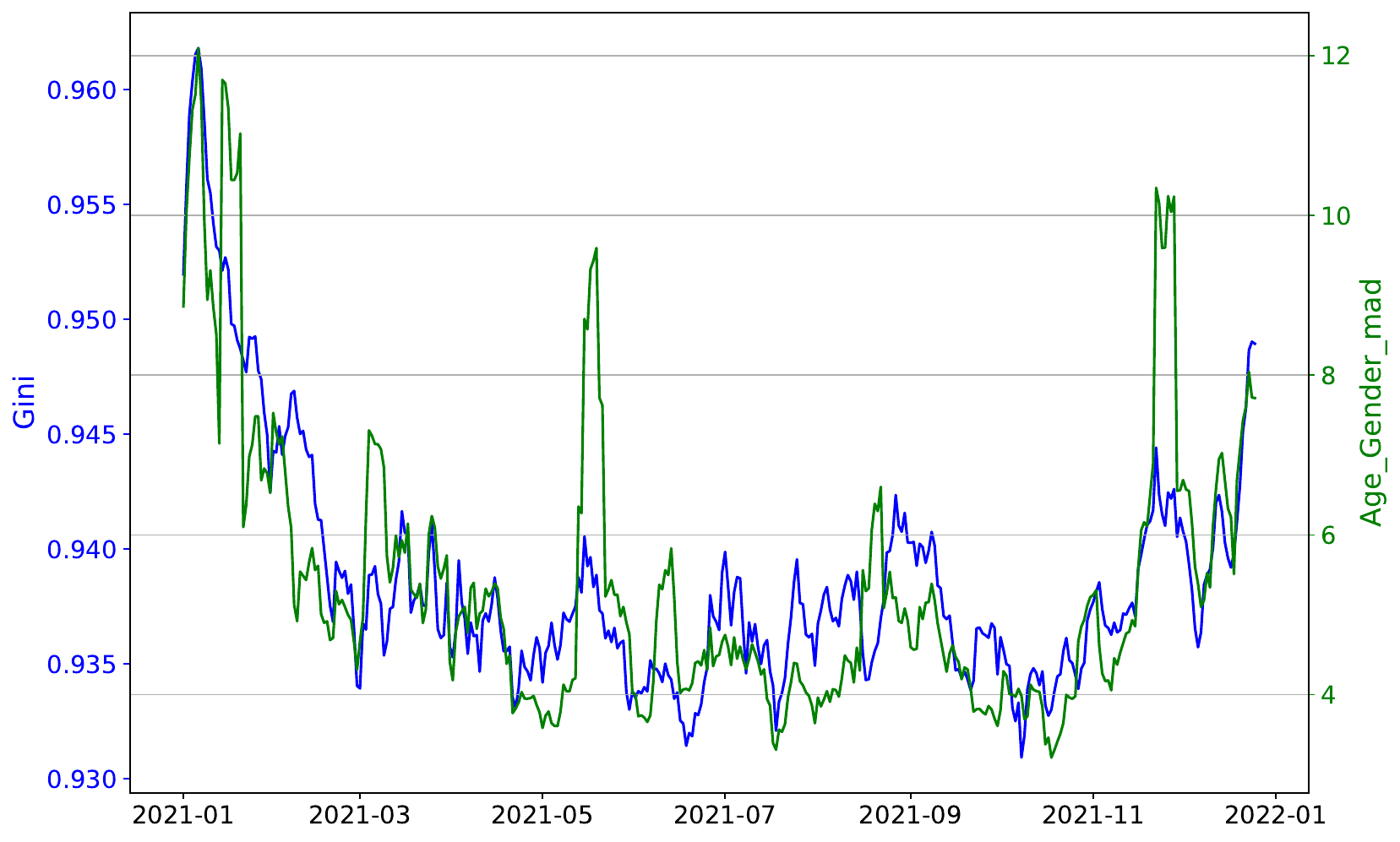}
        \caption{GINI vs. Age \& Gender (MAD)}
        \label{fig:tt2a}
    \end{subfigure}%
    \hfill
    \begin{subfigure}[t]{0.24\textwidth}
        \includegraphics[width=\linewidth]{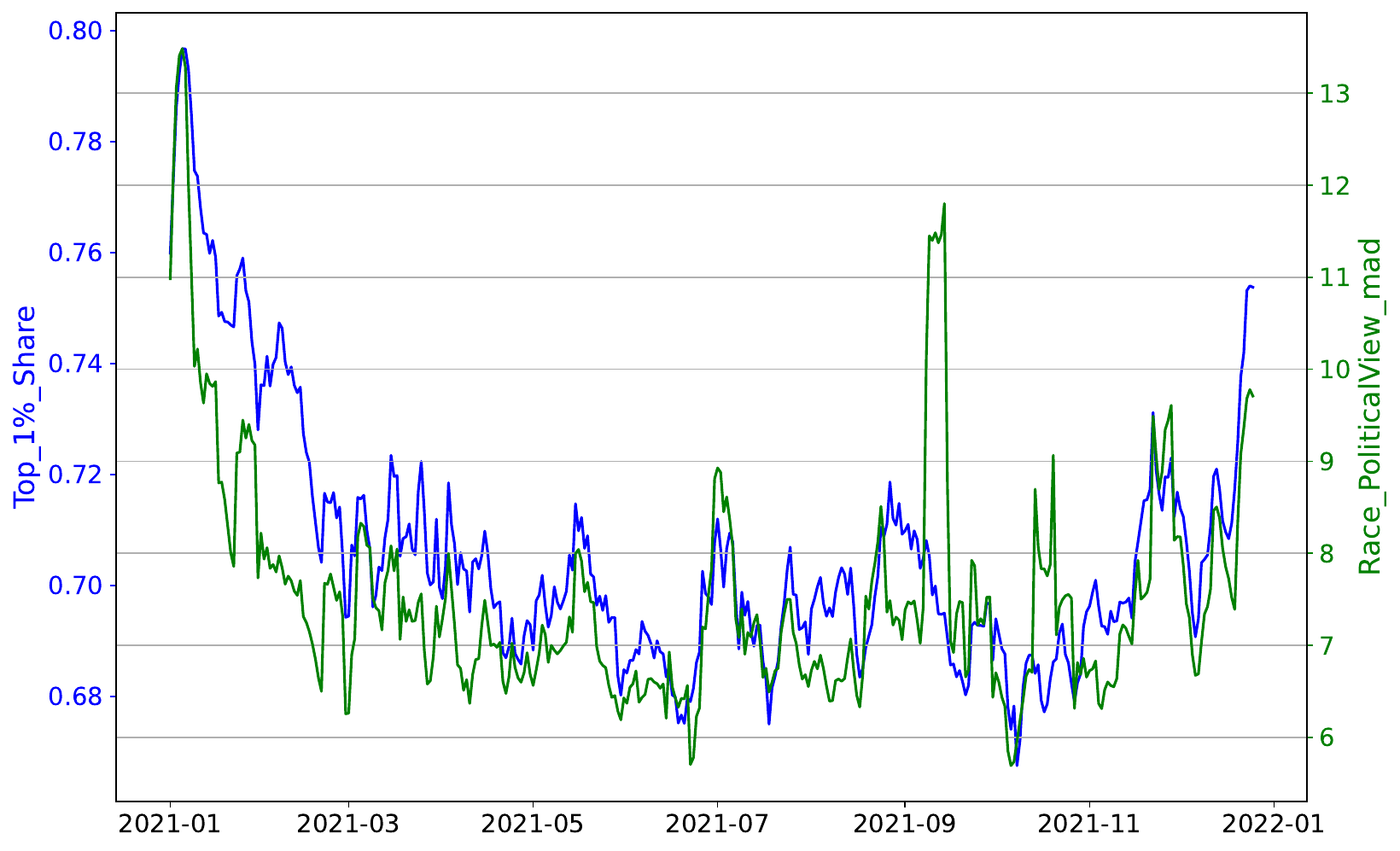}
        \caption{T1PS vs. Race \& Political View (MAD)}
        \label{fig:tt2b}
    \end{subfigure}
    \hfill
    \begin{subfigure}[t]{0.24\textwidth}
        \includegraphics[width=\linewidth]{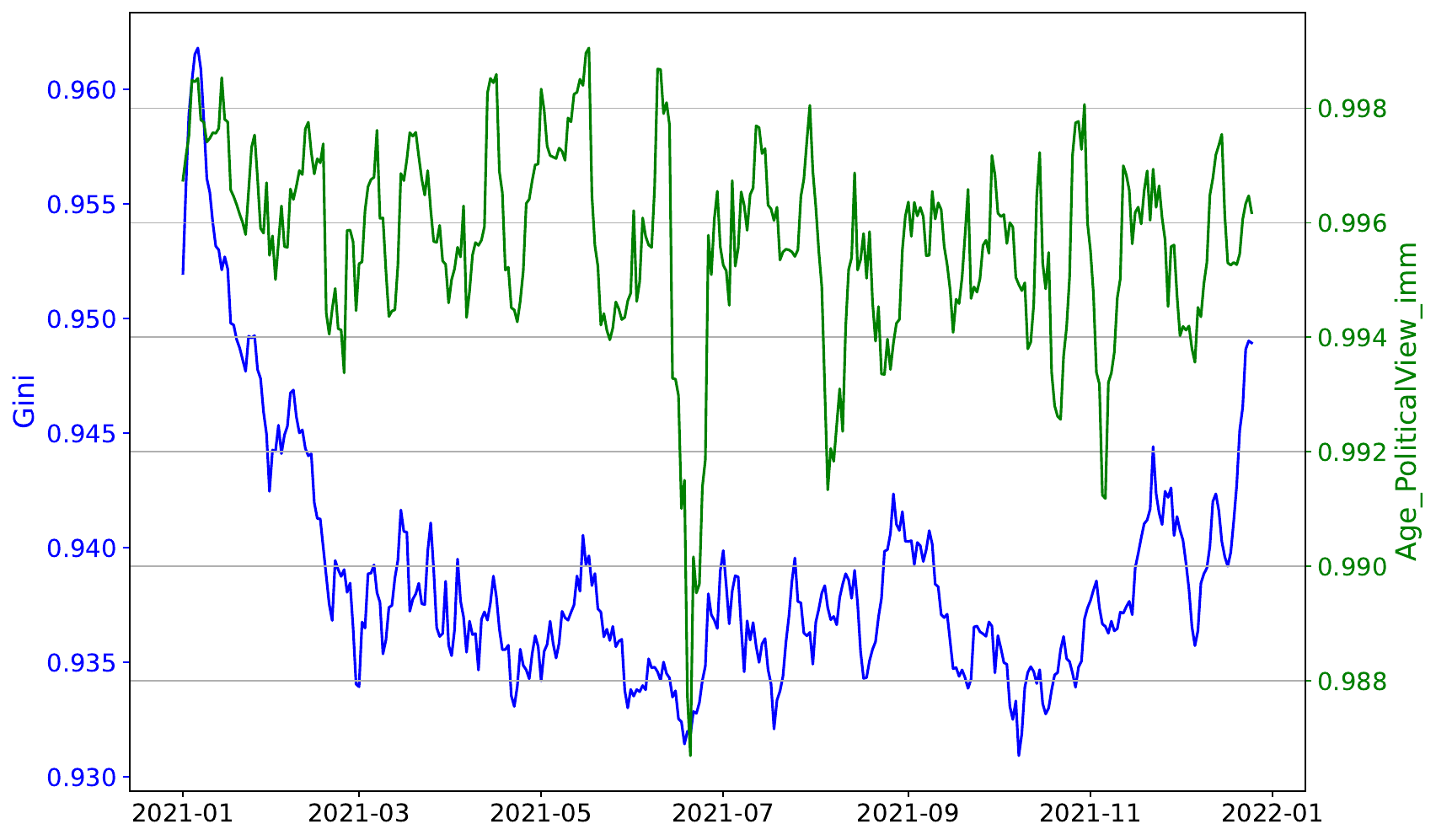}
        \caption{GINI vs. Age \& Political View (IMM)}
        \label{fig:tt2c}
    \end{subfigure}
    \hfill
    \begin{subfigure}[t]{0.24\textwidth}
        \includegraphics[width=\linewidth]{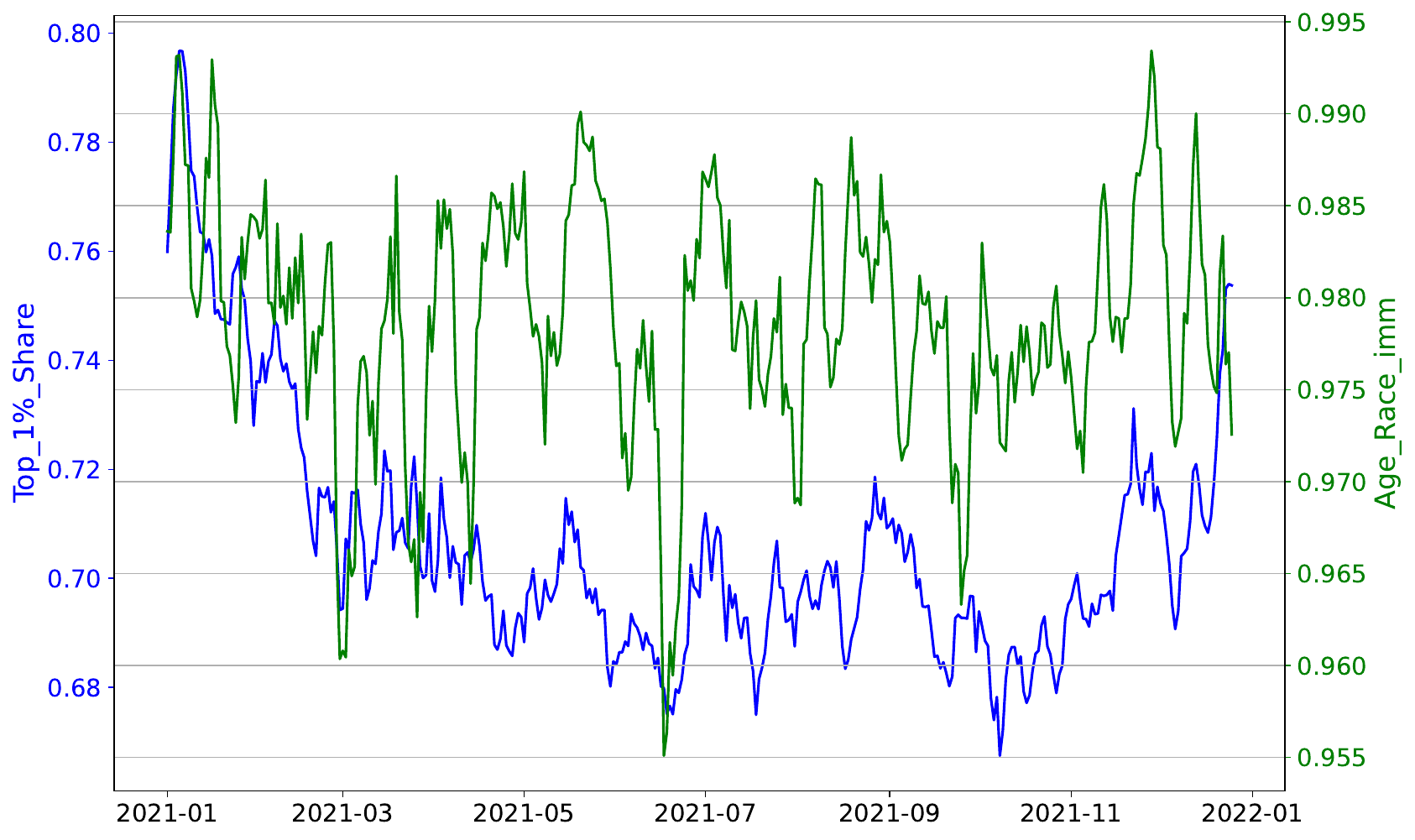}
        \caption{T1PS vs. Age \& Race (IMM)}
        \label{fig:tt2d}
    \end{subfigure}
    \caption{Daily tracking of inequality metrics (blue) and intersectional bias metrics (green) over 2021.}
    \label{fig:timeseriesint}
\end{figure*}

\section{Results} \label{sec:results}

Figure~\ref{fig:marginalsp} shows the correlation between the demographic-free metrics and the demographic disparity metrics applied to each of the demographic variables, along with their statistical significance (p-values). With one exception, all of the demographic bias metrics are positively and significantly correlated with demographic-free metrics. Gini and T1PS have correlation of up to 0.61 with MAD for all demographic variables. We see that IMM for all demographic variables is less correlated with demographic-free measures than MAD is across demographic variables. One variable that stands out is political view. It has a low Spearman's correlation value with both GINI and T1PS, and this value is not statistically significant.

In Figure~\ref{fig:marginalsp}, we examine the marginal demographic attributes (\eg groups defined only by age bucket), whereas in Figure~\ref{fig:intsp} we consider pairwise intersectional attributes (\eg groups defined by both age and gender). Here, we see generally stronger positive correlation (up to 0.78) between the demographic-free inequality measures and MAD across all pairwise intersections. This is suggestive that demographic-free measures may be picking up on inequality across groupings that are not typically thought of in fairness analysis, including those groupings that are less easily measured. Interestingly, once we consider intersectional groups, IMM continues to remain less correlated than MAD with inequality metrics. The low correlation in the presence of political view as an axis of discrimination largely disappears in the intersectional case, with the exception of the combination with age. However, the age plus political view combination also has a non-significant p-value, meaning we cannot conclude from either the marginal or the intersectional case that a positive correlation exists.

Given that we have calculated these correlation values, we now turn our attention to what the actual data for these metrics looks like over the year. Figures~\ref{fig:timeseriesmarg} and \ref{fig:timeseriesint} show a few exemplar pairs of metrics plotted as a time series over days in 2021. We observe that metric pairs with higher Spearman's correlations have tighter time series correspondence---\eg in Figures~\ref{fig:tt1a}, \ref{fig:tt2a} and \ref{fig:tt2b} (correlation values of 0.6, 0.65, and 0.78, respectively)---while pairs with lower correlations have time series plots that appear almost uncorrelated---\eg in Figure~\ref{fig:tt1d} and \ref{fig:tt2c} (correlation values of 0.02 and 0.26, respectively). We omit the remaining time series plots as they follow these same trends.

\section{Conclusion}
\label{sec:conclusion}

In this work, we investigated the empirical relationship between demographic-free measures of inequality and demographic disparities in engagements, using Twitter as a case study. By leveraging a unique dataset that includes Twitter accounts with labeled demographics, we found a positive correlation between demographic-free measures of inequality and measures of demographic disparities (Figure~\ref{fig:spearman_main}). This suggests that in situations where demographic labels are unavailable (\eg the production Twitter system), using inequality metrics for system tuning may also result in lower group-wise disparities across demographic groups. 

Our results also suggest that demographic-free measures of engagement inequality are not correlated with political variables. This is perhaps a positive outcome, as social networks have been accused of censoring content from political conservatives~\cite{PewSocialMediaCensorship2020}. According to our empirical observations, using inequality metrics for system tuning, at least in the case of Twitter, would not cause a significant impact on engagement between users in different political groups.

This paper presents ongoing work exploring the relationship between demographic-free inequality metrics and standard demographic bias metrics. While our findings provide valuable insights, we view this as a starting point for further research and discussion in the field of algorithmic fairness, particularly in contexts where demographic data is limited or unavailable.

\subsection{Limitations and Future Work}
\label{sec:limitations}

Our study, while offering valuable insights, has several limitations that point to directions for future research:

\paragraph{Engagement vs. Impression Inequality} Due to data availability, we focus on engagement inequality rather than impression inequality. While engagement metrics are valuable, they depend on user behavior, adding complexity to our analysis. Future work should aim to analyze impression data, if available, to gain more direct insights into platform content distribution mechanisms.

\paragraph{Dataset Coverage and Generalizability} Our dataset, though large and demographically representative of the U.S. population, doesn't capture the entire Twitter user base. This limitation may affect the generalizability of our findings. Future research should aim to extend this analysis to Twitter's global user base and other social media platforms to test the broader applicability of demographic-free inequality metrics in addressing algorithmic bias.

\paragraph{Influencer Effects and Natural Inequality} Our analysis doesn't distinguish between inequality arising from natural differences in account popularity (e.g., influencers vs. regular users) and inequality stemming from demographic disparities. This could lead to misinterpretation of results, as reductions in overall inequality might not necessarily indicate reductions in unfair treatment. Future work should explore methods to separate these effects, perhaps by controlling for account type or follower count.

\paragraph{Correlational Nature of Findings} Our results suggest a relationship between demographic-free inequality and demographic disparities, but these conclusions are correlational rather than causal. Future studies should conduct causal experiments, such as A/B tests \cite{ddg}, to definitively establish the impact of reducing demographic-free inequality on demographic disparities.

\paragraph{Scope of Analysis} Our study focuses on author-based inequality metrics and excludes users with very low engagement rates. Future research should investigate reader-based diversity metrics, analyzing the diversity of tweet content and authors in users' timelines. Additionally, different approaches to handling outliers and low-engagement users should be explored to understand patterns among users whose tweets receive little to no engagement.

\begin{acks}
The authors thank Alexi Quintana Mathe and the members of the Lazer Lab for providing access to the Twitter Panel dataset.
\end{acks} 

\bibliographystyle{ACM-Reference-Format}
\bibliography{ref}


\end{document}